\documentclass[draft]{agujournal2019}
\usepackage{url} %this package should fix any errors with URLs in refs.
\usepackage{lineno}
\usepackage[inline]{trackchanges} %for better track changes. finalnew option will compile document with changes incorporated.
\usepackage{soul}
\usepackage{amsmath}
\drafttrue
\journalname{Geophysical Research Letters}

\begin{document}

%%%%%%%%%%%%%%%%%%%%%%%%%%%%%%%%%%%%%%%%%%%%%%%
%  TITLE
%
% (A title should be specific, informative, and brief. Use
% abbreviations only if they are defined in the abstract. Titles that
% start with general keywords then specific terms are optimized in
% searches)
%
%%%%%%%%%%%%%%%%%%%%%%%%%%%%%%%%%%%%%%%%%%%%%%%

% Example: \title{This is a test title}

\title{Structure of energy flux densities in Earth's magnetotail reconnection regions: Bridging simulation and observation}

%%%%%%%%%%%%%%%%%%%%%%%%%%%%%%%%%%%%%%%%%%%%%%%
%
%  AUTHORS AND AFFILIATIONS
%
%%%%%%%%%%%%%%%%%%%%%%%%%%%%%%%%%%%%%%%%%%%%%%%

% Authors are individuals who have significantly contributed to the
% research and preparation of the article. Group authors are allowed, if
% each author in the group is separately identified in an appendix.)

% List authors by first name or initial followed by last name and
% separated by commas. Use \affil{} to number affiliations, and
% \thanks{} for author notes.
% Additional author notes should be indicated with \thanks{} (for
% example, for current addresses).

% Example: \authors{A. B. Author\affil{1}\thanks{Current address, Antartica}, B. C. Author\affil{2,3}, and D. E.
% Author\affil{3,4}\thanks{Also funded by Monsanto.}}

\authors{Cara L. Waters\affil{1}\thanks{Now at Queen Mary University of London}, Jonathan P. Eastwood\affil{1}, Harry C. Lewis\affil{1}, Na\"is Fargette\affil{2}, Marit \O ieroset\affil{3}, Tai D. Phan\affil{3}, David L. Newman\affil{4}, and Martin V. Goldman\affil{4}}

% \affiliation{1}{First Affiliation}
% \affiliation{2}{Second Affiliation}
% \affiliation{3}{Third Affiliation}
% \affiliation{4}{Fourth Affiliation}

\affiliation{1}{Department of Physics, Imperial College London, London, United Kingdom}
\affiliation{2}{Institut de Recherche en Astrophysique et Plan\'etologie, CNES, CNRS, Toulouse, France}
\affiliation{3}{Space Sciences Laboratory, University of California, Berkeley, CA, USA}
\affiliation{4}{Centre for Integrated Plasma Studies, University of Colorado, Boulder, CO, USA}
%(repeat as many times as is necessary)

% Corresponding author mailing address and e-mail address:

% (include name and email addresses of the corresponding author.  More
% than one corresponding author is allowed in this LaTeX file and for
% publication; but only one corresponding author is allowed in our
% editorial system.)

% Example: \correspondingauthor{First and Last Name}{email@address.edu}

\correspondingauthor{Cara L. Waters}{c.waters@qmul.ac.uk}

%%%%%%%%%%%%%%%%%%%%%%%%%%%%%%%%%%%%%%%%%%%%%%%
% KEY POINTS
%%%%%%%%%%%%%%%%%%%%%%%%%%%%%%%%%%%%%%%%%%%%%%%
%  List up to three key points (at least one is required)
%  Key Points summarize the main points and conclusions of the article
%  Each must be 140 characters or fewer with no special characters or punctuation and must be complete sentences

% Example:
% \begin{keypoints}
% \item	List up to three key points (at least one is required)
% \item	Key Points summarize the main points and conclusions of the article
% \item	Each must be 140 characters or fewer with no special characters or punctuation and must be complete sentences
% \end{keypoints}

\begin{keypoints}
\item Magnetic reconnection separatrices are identified as a distinct region in a statistical study of magnetotail reconnection for the first time
\item Electron and electromagnetic energy flux densities are largest in the separatrices, exceeding those in the exhaust
\item Ion enthalpy flux density is typically largest in the exhaust, giving a spatially organised partition of energy transport
\end{keypoints}

%%%%%%%%%%%%%%%%%%%%%%%%%%%%%%%%%%%%%%%%%%%%%%%
%
%  ABSTRACT and PLAIN LANGUAGE SUMMARY
%
% A good Abstract will begin with a short description of the problem
% being addressed, briefly describe the new data or analyses, then
% briefly states the main conclusion(s) and how they are supported and
% uncertainties.

% The Plain Language Summary should be written for a broad audience,
% including journalists and the science-interested public, that will not have 
% a background in your field.
%
% A Plain Language Summary is required in GRL, JGR: Planets, JGR: Biogeosciences,
% JGR: Oceans, G-Cubed, Reviews of Geophysics, and JAMES.
% see http://sharingscience.agu.org/creating-plain-language-summary/)
%
%%%%%%%%%%%%%%%%%%%%%%%%%%%%%%%%%%%%%%%%%%%%%%%

%% \begin{abstract} starts the second page

\begin{abstract}
Magnetic reconnection is important in both space and laboratory contexts because it changes the topology of magnetic fields and converts magnetic energy into particle energies. Establishing where this energy conversion occurs is critical for understanding how reconnection partitions energy. Using a novel machine learning approach based on simulations that explicitly identifies the separatrix regions for the first time, the partition of energy flux densities within the magnetic reconnection site is evaluated through statistical study of magnetotail reconnection events observed in situ. The separatrices are found to have the largest electron energy flux densities, roughly twice those in the exhaust, and Poynting flux densities about four times those in the inflows, whereas typical ion enthalpy flux densities are largest in the exhaust. The separatrices are therefore sites of intense electron and electromagnetic energy transport, although their small thickness means that the exhaust may still carry most of the total energy flux.
\end{abstract}

\section*{Plain Language Summary}
Magnetic reconnection is a plasma process in which oppositely directed magnetic field lines break and rejoin, converting stored magnetic energy into the motion and heat of charged particles. It powers solar flares, drives disturbances in near-Earth space, and limits confinement in laboratory fusion devices. A central open question is how the released energy is divided between particle populations, and where within a reconnection site this happens. Spacecraft measurements of reconnection events in Earth’s magnetotail are combined with a machine learning classifier, trained on simulations, that identifies which region of the reconnection site each measurement sampled. This resolves statistically, for the first time, the separatrices: the thin boundary layers dividing field lines that have reconnected from those that have not. Across many events, the electron energy flux density, the rate at which electrons carry energy through a given area, is found to be highest in the separatrices. Earlier observations did not distinguish these layers from their surroundings. The wider exhaust may still carry more energy in total, but electron energy transport is most intense in the separatrices. Since energised electrons stream along the field away from the reconnection site, this may affect how reconnection couples to the wider system.

%%%%%%%%%%%%%%%%%%%%%%%%%%%%%%%%%%%%%%%%%%%%%%%
%
%  BODY TEXT
%
%%%%%%%%%%%%%%%%%%%%%%%%%%%%%%%%%%%%%%%%%%%%%%%

%%% Suggested section heads:
% \section{Introduction}
%
% The main text should start with an introduction. Except for short
% manuscripts (such as comments and replies), the text should be divided
% into sections, each with its own heading.

% Headings should be sentence fragments and do not begin with a
% lowercase letter or number. Examples of good headings are:

% \section{Materials and Methods}
% Here is text on Materials and Methods.
%
% \subsection{A descriptive heading about methods}
% More about Methods.
%
% \section{Data} (Or section title might be a descriptive heading about data)
%
% \section{Results} (Or section title might be a descriptive heading about the
% results)
%
% \section{Conclusions}

\section{Introduction}
%Text here ===>>>
Magnetic reconnection is a ubiquitous plasma process that rapidly transforms magnetic energy into particle energy, driving eruptive events throughout the universe from solar flares and geomagnetic substorms to high-energy emissions from neutron stars and black holes \cite{guo2024}. This fundamental process governs energy release, plasma transport, and confinement, yet it remains unclear how the released energy is partitioned and transported between electromagnetic fields and particle populations across the distinct regions of a reconnection site \cite{liu2025b}. Resolving this is important not only for understanding magnetospheric energy transfer (which can have severe socioeconomic consequences \cite{eastwood2017, oughton2018}), but also for identifying how reconnection redistributes energy and particles in fusion plasmas, where it can degrade confinement and carry energy away from the core \cite{chapman2011}. Another application is in astrophysical plasmas, where only remote sensing is possible and relies on observing electron signatures \cite{uzdensky2011}.

\begin{figure}
    \centering
    \includegraphics[width=0.75\linewidth]{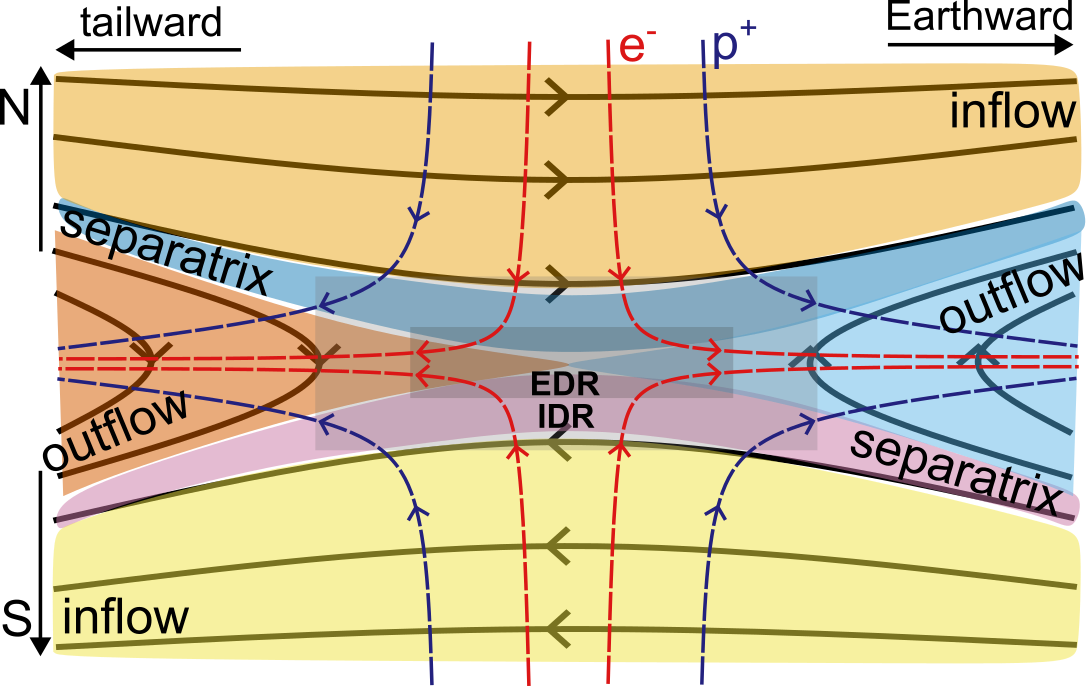}
    \caption{The reconnection site with northern and southern inflow and separatrix regions, and Earthward and tailward outflow regions identified. Magnetic field lines are shown in black solid lines, with proton flow lines in dashed dark blue and electron flow lines in dashed red. The ion and electron diffusion regions (IDR and EDR) are shown as grey shaded regions at the centre. The figure is schematic and not to scale: the EDR and the separatrix layers are much thinner relative to the IDR and the exhaust than shown.}
    \label{fig:rxn}
\end{figure}

Reconnection restructures magnetic topology, accelerating inflowing plasma into high speed exhausts. The separatrices form narrow transition layers separating inflow and outflow domains and contain mixed plasma populations \cite{lapenta2015}. The overall geometry of the reconnection site and the spatial relationship between these regions are illustrated schematically in Figure \ref{fig:rxn}. Plasma transport into the exhaust occurs predominantly along separatrices rather than through the diffusion regions, enabling reconnection to drive large-scale convection despite the small spatial extent of the diffusion region \cite{kulsrud2001}. These layers are dynamically active: they are permeable to energetic ions \cite{holmes2019, phan2022}; host intense parallel electric fields and wave activity \cite{graham2015, jaynes2016}; and support vorticity generation through inflow--outflow interaction \cite{ahmadi2022}. However, the relative roles of separatrices, diffusion regions, and exhausts in mediating energy conversion and transport remain poorly quantified.

Energy conversion during reconnection can be described through conservation of energy densities,

\begin{equation}
\frac{\partial}{\partial t}\bigg(U_{EM} + \sum_sU_s\bigg) + \nabla \cdot \bigg(\mathbf{S} + \sum_s \mathbf{Q}_s\bigg) = 0,
\end{equation}
where $U_\text{EM}$ is the electromagnetic energy density, $U_s$ is the kinetic energy density of particle species $s$, $\mathbf{Q}_s$ is the energy flux density of particle species $s$, and $\mathbf{S} = (\mathbf{E} \times \mathbf{B})/\mu_0$ is the Poynting flux density. The particle energy flux density can be decomposed as

\begin{equation}
\mathbf{Q}_s = \mathbf{K}_s + \mathbf{H}_s + \mathbf{q}_s,
\end{equation}
where $\mathbf{K}_s = \frac12 m_sn_su_s^2\mathbf{u}_s$ is the bulk kinetic energy flux density, $\mathbf{H}_s = (\mathbf{u}_s \text{Tr}(\mathrm{P}_s))/2 + \mathbf{u}_s \cdot \mathrm{P}_s$ is the enthalpy flux density, and $\mathbf{q}_s = \int \frac12m_s|\mathbf{v} - \mathbf{u}_s|^2(\mathbf{v} - \mathbf{u}_s) f_s d^3\mathbf{v}$ is the heat flux density, the third moment of the distribution function $f_s$ in the bulk rest frame, which is non-zero only when $f_s$ deviates from a Maxwellian (for example, when it is skewed).

Observations show that electron and ion bulk heating scale linearly with the available magnetic energy per particle, $m_iV_A^2$, where $V_A$ is the Alfv\'en speed based on the upstream (asymptotic) reconnecting magnetic field and density. At the magnetopause $\Delta T_e \approx 0.017 m_iV_A^2$ and $\Delta T_i \approx 0.13 m_iV_A^2$, so ions typically receive the larger share \cite{phan2013, phan2014, oieroset2023}. Magnetotail measurements suggest this relation breaks down at high available magnetic energy \cite{oieroset2024}, consistent with theory predicting that only the total particle heating scales linearly with available magnetic energy \cite{haggerty2015}. Energy flux studies have largely focused on reconnection inflows and outflows, where ion enthalpy flux density dominates the outflowing energy on average for symmetric collisionless magnetotail reconnection, although localised Poynting flux enhancements occur within the ion diffusion region \cite{eastwood2013}. Significant electron energy flux densities are observed near the electron diffusion region \cite{eastwood2020, fargette2024}, with electron-dominated transport at electron scales transitioning to ion dominance at ion inertial length scales \cite{fargette2024}. Guide fields further modify energy transport by enabling Poynting flux to be carried from inflow to outflow with limited conversion \cite{birn2010}.

A key unresolved problem is therefore to determine where energy conversion and transport primarily occur within the reconnection system, and how the partition between fields, electrons, and ions varies between inflow, separatrix, and outflow regions. Recent machine learning approaches have enabled automated identification of these regions in particle-in-cell simulations \cite{waters2024} and spacecraft observations \cite{waters2025}. Here we use such classifications to perform a statistical investigation of energy partition across reconnection regions in Earth’s magnetotail, providing constraints on how reconnection redistributes energy in collisionless plasmas. The separatrices are explicitly identified, allowing the energy partition within them to be determined. Overall, the separatrices are found to have the largest electron and electromagnetic energy flux densities in the reconnection site, whereas typical ion energy flux densities are largest in the outflows. This builds on previous studies using the Cluster mission \cite{eastwood2013, tyler2016}, where the separatrices and electron-scale physics could not be resolved.

\section{Data and Methods}

This study analysed anti-parallel magnetic reconnection events observed by the Magnetospheric Multiscale (MMS) mission in Earth's magnetotail during the second mission phase (2017--2020). For intervals after 7 June 2018, MMS4 data were excluded due to the failure of one electron spectrometer pair. Magnetic field measurements from the fluxgate magnetometer (FGM) \cite{russell2016} were used at 128 Hz resolution. Plasma moments were obtained from the fast plasma investigation (FPI) \cite{pollock2016}, with burst mode resolutions of 30 ms for electrons and 150 ms for ions. Electric field measurements from the electric field double probes (EDP) \cite{ergun2016, lindqvist2016} were used at 30 ms resolution.

Events were selected from a catalogue of 26 previously identified magnetotail reconnection events \cite{rogers2023}, based on data availability, measurement quality, and spacecraft trajectories sampling multiple reconnection regions, including inflow, separatrix, and exhaust crossings. Some intervals exhibited overestimation of ambient electron populations, producing unphysical plasma moments such as vanishing number density \cite{liu2025}. To enable a consistent automated statistical analysis, only events without such artefacts were retained, resulting in a final sample of ten reconnection events (Table \ref{tab:events}).

\begin{table}
    \centering
    \caption{The events included in the energy flux analysis, with their asymptotic inflow magnetic field $B_0$, peak plasma number density in the current sheet $n_0$, normalised guide field $B_G/B_0$, and distance downtail in the $X_{\mathrm{GSM}}$ direction.}
    \begin{tabular}{cccccc}
        \textbf{Start} & \textbf{End} & $B_0$ (nT) & $n_0$ (cm$^{-3}$) & $B_G/B_0$ & $X_{\mathrm{GSM}}$ ($R_E$) \\ \hline
        2017-05-28/03:58:30 & 03:59:15 & 21.3 & 0.140 & 0.1 & -19.3 \\
        2017-07-06/15:35:10 & 15:36:00 & 16.1 & 0.439 & 0.1 & -24.1 \\
        2017-07-06/15:46:35 & 15:46:52 & 19.1 & 0.199 & 0.2 & -24.2 \\
        2017-07-11/22:33:30 & 22:34:15 & 7.81 & 0.199 & 0.1 & -21.5 \\
        2017-07-17/07:48:45 & 07:49:30 & 16.4 & 0.173 & 0.1 & -18.1 \\
        2018-08-27/12:15:30 & 12:16:00 & 16.1 & 0.159 & 0.1 & -21.1 \\
        2018-09-10/17:14:45 & 17:15:30 & 29.1 & 2.76 & 0.2 & -18.2 \\
        2019-07-25/21:41:10 & 21:41:55 & 10.3 & 0.204 & 0.1 & -24.0 \\
        2019-08-31/12:03:00 & 12:03:30 & 32.4 & 0.710 & 0.1 & -13.0 \\
        2020-08-03/01:05:40 & 01:06:20 & 20.1 & 0.188 & 0.1 & -27.5 \\
    \end{tabular}
    \label{tab:events}
\end{table}

To allow inter-event comparison, all quantities were normalised using characteristic upstream conditions. For each event, the asymptotic reconnecting magnetic field $B_0$ was determined by taking the median value of $B_x$ over an inflow period determined by eye, and the peak current sheet density $n_0$ was taken as the maximum ion number density measured at the current sheet centre. The ion Alfv\'en speed $V_{Ai0} = B_0/\sqrt{\mu_0m_in_0}$ was then calculated from these values. Physical quantities were then expressed in dimensionless form: magnetic field $\mathbf{B}' = \mathbf{B}/B_0$, number density $n_s' = n_s/n_0$, electric field $\mathbf{E}' = \mathbf{E}/(B_0V_{Ai0})$, bulk velocity $\mathbf{V}_s' = \mathbf{V}_s/V_{Ai0}$, and energy flux density $\mathbf{Q}_s' = (\mu_0/(V_{Ai0}B_0^2))\mathbf{Q}_s$. The guide field $B_G/B_0$ was estimated by eye from the normalised time series within the inflow regions of each event. This could not be taken as an average due to the varying trajectories through each event and different coverage of each side of the current sheet. Across the event set, $B_G/B_0$ is consistently within the range of 0.1--0.2 (Table \ref{tab:events}).

\begin{figure}
    \centering
    \includegraphics[width=0.75\linewidth]{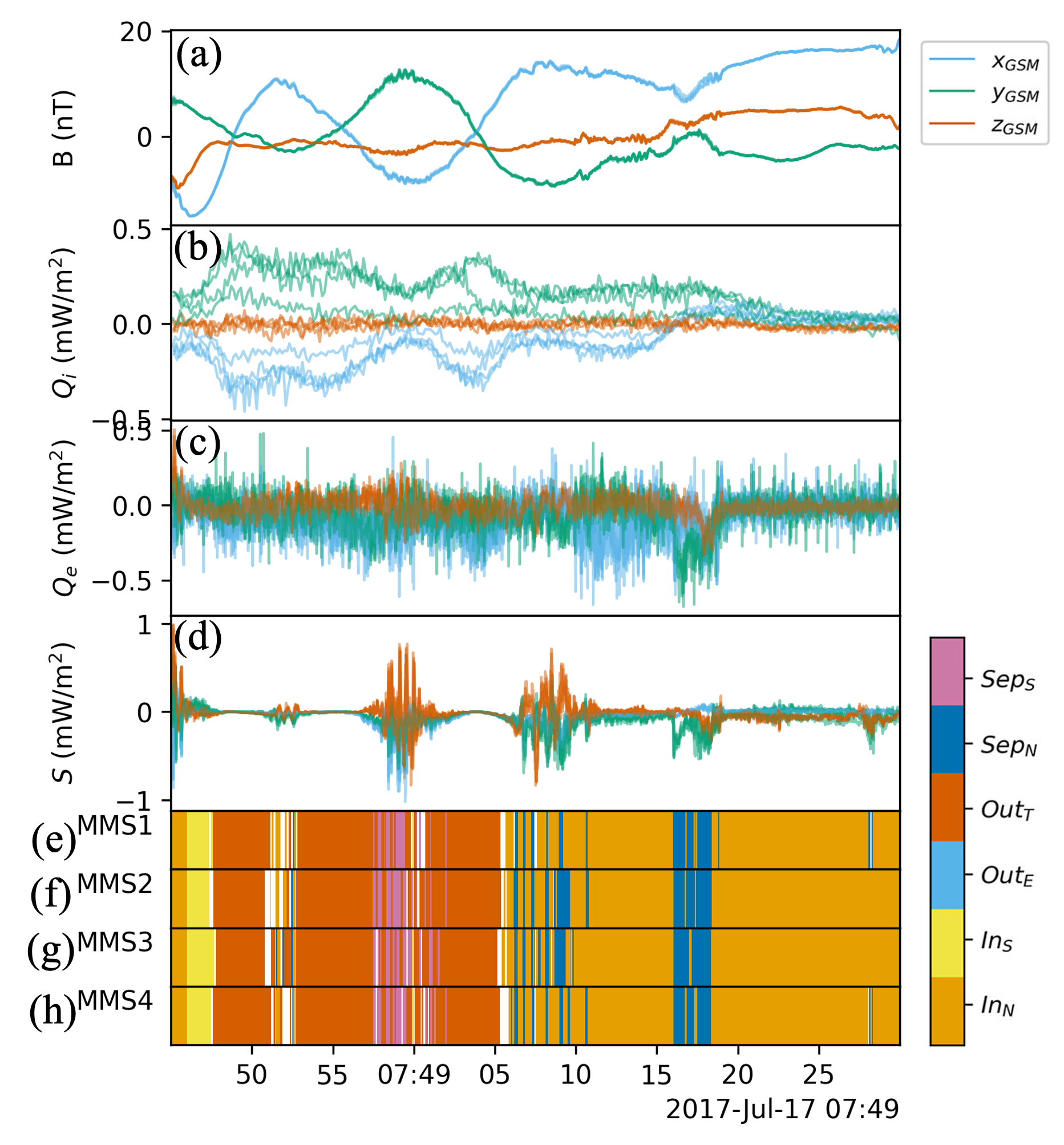}
    \caption{A reconnection event with region classifications for each of the four MMS spacecraft, showing (a) magnetic field $\mathbf{B}$, (b) ion energy flux density $\mathbf{Q}_i$, (c) electron energy flux density $\mathbf{Q}_e$, (d) Poynting flux density $\mathbf{S}$, and (e--h) region classifications, where white regions indicate a classification with less than 90\% certainty. Time series for all four of the spacecraft are overlaid.}
    \label{fig:timeseries}
\end{figure}

Spatial regions within each event were identified using a data-driven classification approach. A particle-in-cell simulation of symmetric reconnection with a guide field of $0.1 B_0$, representative of plasma sheet conditions, was first segmented into inflow (northern $In_N$ and southern $In_S$), separatrix (northern $Sep_N$ and southern $Sep_S$), and outflow (Earthward $Out_E$ and tailward $Out_T$) regions using \textit{k}-means clustering \cite{waters2024}. These labelled simulation trajectories were then used to train a recurrent neural network to recognise temporal signatures associated with each region. The trained network was validated against a well constrained multi-spacecraft reconnection event before being applied to the full event set to obtain region classifications along spacecraft trajectories \cite{waters2025}. Figure \ref{fig:timeseries} shows the resulting classifications for a representative event. This event of 17 July 2017 shows the plasma sheet flapping in the tail, with MMS going between both inflow regions as a result. As the X-line retreats tailward over the spacecraft, this motion causes MMS to cross from the tailward outflow to the northern inflow, with spikes in the Poynting flux density $\mathbf{S}$ being seen, corresponding to crossing the separatrices.

\section{Comparison of Particle and Field Energy Flux Densities}

For each point in the time series, the normalised particle and electromagnetic energy flux densities were calculated. These were attributed to the regions according to their classification found by the machine learning algorithm. A statistical picture was then built up across all ten events, with good coverage of each of the regions obtained \cite{waters2025}.

\begin{figure*}
    \centering
    \includegraphics[width=\linewidth]{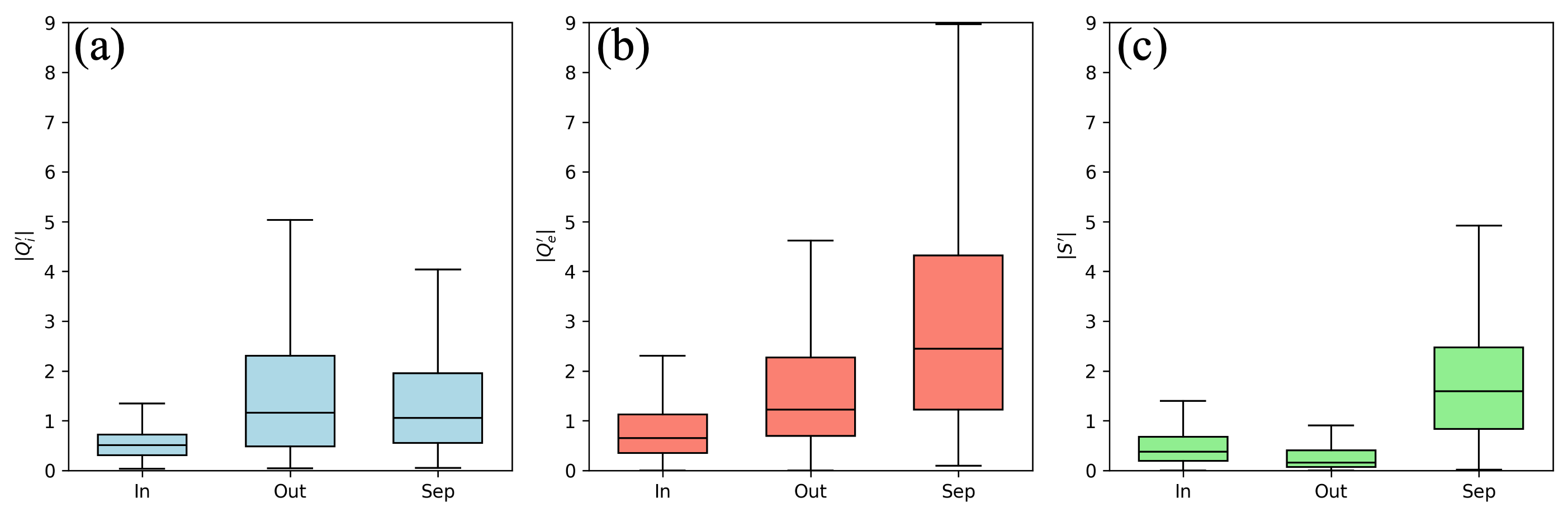}
    \caption{Box plots showing the distribution of the magnitude of the total ion and electron normalised energy flux densities and the Poynting flux density across the pairs of the regions identified, with (a) magnitude of normalised ion energy flux density $|\mathbf{Q}_i'|$, (b) electron energy flux density $|\mathbf{Q}_e'|$, and (c) Poynting flux density $|\mathbf{S}'|$. Whiskers are positioned at 1.5 times the interquartile range from the quartiles, and the upper quartile, median, and lower quartile are shown. Outliers from this range are omitted.}
    \label{fig:mags}
\end{figure*}

Figure \ref{fig:mags} shows the distributions of total particle energy flux densities and Poynting flux density across inflow, separatrix, and outflow regions for all ten events combined. Ion energy transport (Figure \ref{fig:mags}a) is weakest in the inflows. In contrast, the separatrices and outflows both exhibit substantially larger ion energy flux densities, with similar median values but broader variability in the outflows, indicating intermittent enhancement of ion energy transport within structured exhaust regions.

Electron energy transport (Figure \ref{fig:mags}b) shows a clearer regional ordering. On average, separatrices exhibit the largest electron energy flux densities, followed by outflows, with inflows contributing least. Overlap between the outflow distributions and those of both inflows and separatrices suggests that localised structures within exhausts can produce electron energy transport comparable to that observed at separatrices. Across all regions, electron energy flux densities can reach magnitudes comparable to those of ions.

The Poynting flux density (Figure \ref{fig:mags}c) is most prominent in the separatrices, where its typical magnitude exceeds that in the outflows by more than an order of magnitude. Although extreme outflow values occasionally approach separatrix levels, electromagnetic energy transport is generally much weaker in the exhaust. Within separatrices, the Poynting flux density is comparable on average to the electron energy flux density, whereas in inflows it is similar to the ion energy flux density and in outflows it is typically smaller than the ion contribution. These trends differ from simulation results \cite{waters2024}, where separatrix Poynting flux was found to exceed electron enthalpy flux by an order of magnitude and inflow Poynting flux dominated ion energy transport.

Flux densities alone do not determine whether separatrix energy transport results from local energy conversion or from energy carried into the separatrices from elsewhere, for example by electrons accelerated near the X-line streaming along the separatrices. Distinguishing these requires the local energy conversion rate $\mathbf{J} \cdot \mathbf{E}'$ or the divergence of the energy flux within the separatrix layers, which cannot be reliably determined across the present event set.

\section{Energy Partition Within Species}

\begin{table*}
\centering
\caption{Energy flux density magnitudes normalised by inflowing normalised Poynting flux $S'_\textrm{in}$ for each of the three regions, giving the median, interquartile range (IQR), and 90th percentile (P90). These are for the ion kinetic energy flux density $|\mathbf{K}'_i|$, ion enthalpy flux density $|\mathbf{H}'_i|$, ion heat flux density $|\mathbf{q}'_i|$, total ion energy flux density $|\mathbf{Q}'_i|$, electron kinetic energy flux density $|\mathbf{K}'_e|$, electron enthalpy flux density $|\mathbf{H}'_e|$, electron heat flux density $|\mathbf{q}'_e|$, and Poynting flux density $|\mathbf{S}'|$, with the number of measurements within each region $N$.}
\begin{tabular}{ccccccccccc}
Quantity&Statistic &       In &      Out &     Sep & \hspace{1em} & Quantity & Statistic & In & Out & Sep \\ \hline
$|\mathbf{K}'_i|$   &Median    &    0.036 &    0.073 &   0.029 & & $|\mathbf{K}'_e|$   &Median    &    0.001 &    0.003 &   0.023 \\
        &IQR       &    0.069 &    0.191 &   0.090 & & & IQR       &    0.004 &    0.010 &   0.107 \\
        &P90       &    0.139 &    0.609 &   0.514 & & & P90       &    0.013 &    0.034 &   0.823 \\
$|\mathbf{H}'_i|$   &Median    &    0.650 &    2.045 &   1.201 & & $|\mathbf{H}'_e|$   &Median    &    0.754 &    1.937 &   4.252 \\
        &IQR       &    0.775 &    3.460 &   2.191 & & & IQR       &    1.193 &    2.798 &   5.249 \\
        &P90       &    2.172 &    8.515 &  10.103 & & & P90       &    2.745 &    6.030 &  10.434 \\
$|\mathbf{q}'_i|$   &Median    &    0.550 &    0.783 &   1.134 & & $|\mathbf{q}'_e|$   &Median    &    0.721 &    1.107 &   1.739 \\
        &IQR       &    0.633 &    1.301 &   1.428 & & & IQR       &    0.941 &    1.374 &   2.546 \\
        &P90       &    1.330 &    2.900 &   2.894 & & & P90       &    2.293 &    3.927 &   6.651 \\
$|\mathbf{Q}'_i|$   &Median    &    0.993 &    2.898 &   2.293 & & $|\mathbf{Q}'_e|$   &Median    &    1.116 &    2.257 &   4.455 \\
        &IQR       &    1.007 &    4.598 &   3.798 & & & IQR       &    1.519 &    3.216 &   5.610 \\
        &P90       &    2.843 &    9.732 &  10.229 & & & P90       &    3.824 &    7.438 &  14.180 \\ \\
$|\mathbf{S}'|$     &Median    &    1.000 &    0.412 &   4.298 \\
        &IQR       &    1.269 &    0.880 &   4.244 \\
        &P90       &    3.279 &    1.958 &   9.937 \\ \\
        &N         &    12298 &    22735 &    4517 \\ 
\end{tabular}
\label{tab:stats}
\end{table*}

Table \ref{tab:stats} quantifies the partition of particle and electromagnetic energy flux densities across reconnection regions. Ion energy transport is dominated by enthalpy flux in all regions. The median ion enthalpy flux increases from $0.65|\mathbf{S}'_{\mathrm{in}}|$ in the inflows to $2.05|\mathbf{S}'_{\mathrm{in}}|$ in the outflows and $1.20|\mathbf{S}'_{\mathrm{in}}|$ in the separatrices, so typical ion energy transport is strongest in the exhaust. The upper percentiles show a different ordering: the 90th percentile is $8.5|\mathbf{S}'_{\mathrm{in}}|$ in the outflows and exceeds $10|\mathbf{S}'_{\mathrm{in}}|$ in the separatrices. The most intense ion enthalpy flux densities therefore occur in the separatrices, consistent with these layers being permeable to energetic ions \cite{holmes2019, phan2022}. Ion heat flux contributes at a comparable level in the inflows ($0.55|\mathbf{S}'_{\mathrm{in}}|$ median) but becomes secondary to enthalpy transport downstream. Ion bulk kinetic energy flux is small throughout (median $\le 0.07|\mathbf{S}'_{\mathrm{in}}|$), although the broad interquartile range in the outflows reflects intermittent enhancements associated with structured exhaust dynamics. As a result, the total ion energy flux increases by approximately a factor of three from inflow to outflow.

The ion energy flux density in the inflows is comparable to the Poynting flux density (median $|\mathbf{Q}'_i| = 0.99|\mathbf{S}'_\textrm{in}|$). This differs from the low $\beta$ of the magnetotail lobes, where the inflowing energy is almost entirely electromagnetic \cite{eastwood2013}. However, the inflow intervals identified here mostly sample the outer plasma sheet and plasma sheet boundary layer adjacent to the separatrices rather than the lobes, where $\beta$ is substantially higher. The flux magnitudes also include out of plane and field-aligned ion flows, which do not transport energy towards the X-line. The increase in ion energy flux density from inflow to outflow therefore partly reflects thermal energy already present in the inflowing plasma.

Electron energy transport exhibits a distinct regional dependence. Median electron enthalpy flux rises from $0.75|\mathbf{S}'_{\mathrm{in}}|$ in the inflows to $1.94|\mathbf{S}'_{\mathrm{in}}|$ in the outflows and $4.25|\mathbf{S}'_{\mathrm{in}}|$ in the separatrices, while median electron heat flux increases from $0.72|\mathbf{S}'_{\mathrm{in}}|$ to $1.11|\mathbf{S}'_{\mathrm{in}}|$ and $1.74|\mathbf{S}'_{\mathrm{in}}|$ across the same regions. Consequently, the total electron energy flux is largest in the separatrices (median $4.46|\mathbf{S}'_{\mathrm{in}}|$), exceeding the outflow median by roughly a factor of two and the inflow median by a factor of four. Electron bulk kinetic energy flux is negligible in the inflows and outflows (median $\le 0.003|\mathbf{S}'_{\mathrm{in}}|$) but becomes significant in the separatrices, where the 90th percentile reaches $0.82|\mathbf{S}'_{\mathrm{in}}|$. This shows that fast electron flows are present within the separatrix layers. Bulk kinetic energy flux alone cannot determine whether these electrons are accelerated locally or closer to the X-line before streaming along the separatrices. However, it is consistent with the electron acceleration previously observed at magnetotail separatrices \cite{norgren2020}, and the enhanced Poynting flux density in the separatrices (below) requires enhanced electric fields there. These trends demonstrate that electron energy flux density is concentrated in the separatrix layers.

Electromagnetic energy transport shows a complementary spatial structure. The Poynting flux density is largest in the separatrices (median $4.30|\mathbf{S}'_{\mathrm{in}}|$), a factor of $\sim$4 larger than in the inflows and more than ten times larger than in the outflows (median $0.41|\mathbf{S}'_{\mathrm{in}}|$). The upper percentile values further emphasise this contrast, with separatrix Poynting flux density approaching $10|\mathbf{S}'_{\textrm{in}}|$.

Taken together, these results indicate systematic redistribution of energy across the reconnection system. Electromagnetic energy entering through the inflows is primarily converted into ion enthalpy transport in the exhaust, while the separatrices host the largest electron and electromagnetic flux densities. The comparable magnitudes of electron and electromagnetic energy fluxes in separatrix regions suggest that these layers play a central role in mediating cross-scale energy transfer within collisionless reconnection.

\section{Discussion}

This statistical study demonstrates a strong regional asymmetry in the partition of energy transport during magnetotail reconnection. Within the outflows, ion energy transport dominates the particle contribution, with ions carrying approximately 56\% of the total particle energy flux (median $|\mathbf{Q}_i'| = 2.90|\mathbf{S}'_{\mathrm{in}}|$ compared to $|\mathbf{Q}_e'| = 2.26|\mathbf{S}'_{\mathrm{in}}|$). In contrast, the separatrices are characterised by electron-dominated particle energy transport, where electrons carry approximately 66\% of the particle energy flux (median $|\mathbf{Q}_e'| = 4.46|\mathbf{S}'_{\mathrm{in}}|$ compared to $|\mathbf{Q}_i'| = 2.29|\mathbf{S}'_{\mathrm{in}}|$). These results indicate that ions carry most of the particle energy flux density in the exhaust, whereas electrons carry most of it in the separatrix layers.

These results can be compared with previous estimates of energy partition. In Cluster observations of magnetotail reconnection outflows \cite{eastwood2013}, ion enthalpy flux dominated, with electron enthalpy flux around 40\%, Poynting flux 16--42\% and ion bulk kinetic energy flux less than 20\% of the ion enthalpy flux. In the outflows here, the corresponding median ratios are 0.95, 0.20 and 0.04. The smaller ion bulk kinetic contribution is expected, since \citeA{eastwood2013} selected intervals of fast outflow ($|v_{i,x}| \approx 0.45 V_A$) whereas all outflow-classified points are included here. The larger electron contribution may reflect electron-scale structure, such as Hall currents and electron jets, that MMS resolves but Cluster did not. In laboratory experiments and simulations of symmetric anti-parallel reconnection, ions gain 61--64\% of the energy contributed to particles, and around 45\% of the incoming electromagnetic energy leaves as Poynting flux \cite{yamada2014, ji2023}. In those studies the outgoing energy includes both the exhaust and the separatrices. Here, ions carry 56\% of the particle energy flux density in the outflows but only 34\% in the separatrices. Poynting flux accounts for 39\% of the total energy flux density in the separatrices but only 7\% in the outflows. This suggests that much of the outgoing electromagnetic energy in those inventories may be carried in the separatrix region. Because those studies integrate energy gain over a volume, whereas flux densities are reported here, these comparisons are approximate.

The unexpectedly large electron energy transport in the separatrices contrasts with expectations from heating studies suggesting that ions receive the larger share of released magnetic energy in reconnection as a result of the outflows \cite{oieroset2024, phan2014}. However, separatrices are known to host strong electric fields and wave activity capable of accelerating electrons efficiently \cite{oka2022, jaynes2016, graham2015}. Enhanced electron heating and energy fluxes near the electron diffusion region and along separatrix-associated electron outflows \cite{eastwood2020, fargette2024} are consistent with the large electron energy transport identified here. Acceleration of electrons to several keV within separatrix structures \cite{norgren2020} further supports the interpretation that these regions are active sites of energy conversion rather than passive boundaries. Electron energy transport in the separatrices is therefore more intense than previously recognised, and these layers may provide significant signatures relevant for remote sensing in astrophysical \cite{uzdensky2011, sironi2021} and laboratory \cite{law2020, valenzuela2024} contexts.

In contrast, ion energy transport becomes increasingly important downstream. The median ion enthalpy flux in the outflows ($2.05|\mathbf{S}'_{\mathrm{in}}|$) exceeds that in the separatrices ($1.20|\mathbf{S}'_{\mathrm{in}}|$) and inflows ($0.65|\mathbf{S}'_{\mathrm{in}}|$), indicating that reconnection redistributes magnetic energy into ion pressure and bulk flow on larger spatial scales. Although the upper-percentile ion energy fluxes in separatrices can match or exceed those in the exhaust, the typical partition of energy flux density in the exhaust is ion-dominated. Differences in the relative importance of ion heat flux and enthalpy flux between separatrices and outflows imply that distinct energisation mechanisms operate in these regions, with structured exhaust dynamics enhancing ion pressure-driven transport.

Electromagnetic energy transport exhibits a complementary spatial organisation. The Poynting flux density is largest in the separatrices (median $4.30|\mathbf{S}'_{\mathrm{in}}|$), exceeding the outflow median ($0.41|\mathbf{S}'_{\mathrm{in}}|$) by more than an order of magnitude and remaining comparable to particle energy transport in the inflows. This behaviour is consistent with reconnection theory, in which magnetic energy enters through the inflows and is redistributed across the reconnection system \cite{priest2000}. The separatrix Poynting flux density exceeds the inflow value by a factor of approximately 4. This is consistent with the Hall magnetic and electric fields, which are strongest near the separatrices, redirecting the inflowing Poynting flux away from the X-line and across the separatrices into the exhaust \cite{liu2022, payne2024}. The same enhanced electric field structures that supply this additional $\mathbf{E} \times \mathbf{B}$ Poynting flux could also accelerate electrons. This would link the enhanced electromagnetic and electron energy flux densities found in the separatrices, although testing the link directly requires the local energy conversion rate $\mathbf{J}\cdot\mathbf{E}'$ within the separatrix layers. Separatrices also host large amplitude waves and electrostatic structures \cite{graham2015, jaynes2016, norgren2020}, and kinetic Alfv\'en waves propagating along them can carry significant Poynting flux \cite{shay2011}, which may contribute further to the enhanced separatrix Poynting flux density. The reduced Poynting flux density in the exhaust indicates efficient conversion to particle energy. Variations in guide field strength may further influence this partition by allowing a greater fraction of electromagnetic energy to be transported downstream with limited conversion \cite{birn2010}.

Compared with particle-in-cell simulations \cite{waters2024}, the separatrices play a larger role in electron energy transport, and a smaller role in electromagnetic and ion energy transport, than predicted. Since separatrices carry kinetic Alfv\'en waves whose Poynting flux may drive auroral signatures \cite{shay2011}, this raises questions about their role in the wider effects of reconnection. Expanding the event database and combining observations with simulation-based reconstructions will be needed to establish how universally these trends apply and how they depend on upstream conditions.

The results presented here concern energy flux densities. The total energy flux carried by each region also depends on its thickness, which cannot be determined from the spacecraft trajectories in this event set. The separatrix layers are thin, on electron to sub-ion scales, whereas the exhaust broadens to several ion inertial lengths downstream of the X-line. Each exhaust is bounded by two separatrices, so the median values in Table \ref{tab:stats} imply that the exhaust carries the larger total electron energy flux wherever it is more than about four times thicker than a single separatrix layer. This is expected away from the diffusion region. The separatrix regions identified by the classifier are also broadest near the X-line and thin with distance downstream \cite{waters2024}. The separatrices therefore concentrate electron and electromagnetic energy transport, and are likely to contribute significantly to the total near the X-line, while the exhaust is likely to carry most of the total energy flux further downstream. Quantifying these contributions would require multi-spacecraft estimates of layer thickness across a much larger catalogue of events than is currently available.

Overall, the results indicate that reconnection redistributes energy through a spatially organised pathway: electromagnetic energy enters through the inflows, is concentrated into intense electron and electromagnetic energy flux densities in the separatrices, and is carried downstream primarily by ion enthalpy flux in the exhaust. This structured energy flow has important implications for understanding cross-scale coupling in collisionless plasmas, including energy release in planetary magnetospheres, laboratory fusion devices, and astrophysical reconnection systems.

%%%%%%%%%%%%%%%%%%%%%%%%%%%%%%%%%%%%%%%%%%%%%%%
%
% DATA SECTION and ACKNOWLEDGMENTS
%
%%%%%%%%%%%%%%%%%%%%%%%%%%%%%%%%%%%%%%%%%%%%%%%

\section*{Open Research Section}
The recurrent neural network, along with instructions to utilize it and data generated for the events studied here, can be found at \url{https://github.com/carawaters/mms_rnn_run} \cite{Waters_mms_rnn_run_2025}. Simulation data used to obtain region classifications is available in the \textit{Zenodo} data repository \url{https://doi.org/10.5281/zenodo.12570878} \cite{data}.

Data from the Magnetospheric Multiscale (MMS) mission \cite{burch2016} were used in the creation of this manuscript. MMS data are available at \url{https://lasp.colorado.edu/mms/sdc/public} and \url{https://spdf.gsfc.nasa.gov/pub/data/mms/}. We use burst mode magnetic field data from FGM on all four spacecraft \cite{mms1fgm, mms2fgm, mms3fgm, mms4fgm}, burst mode electron moments \cite{mms1fpielec, mms2fpielec, mms3fpielec, mms4fpielec} and ion moments \cite{mms1fpiion, mms2fpiion, mms3fpiion, mms4fpiion} from FPI on all four spacecraft, and burst and fast mode electric field data from EDP on all four spacecraft \cite{mms1edpburst, mms2edpburst, mms3edpburst, mms4edpburst, mms1edpfast, mms2edpfast, mms3edpfast, mms4edpfast}. The MMS data are archived in the NASA Common Data Format (CDF) and so can be plotted using a number of different data display software packages that parse CDF files. Here, the data analysis was performed using PySPEDAS, an implementation of the Space Physics Environment Data Analysis Software (SPEDAS) framework in Python \cite{grimes2022} (\url{https://pyspedas.readthedocs.io/en/latest/}), version 1.5.6 released 20th February 2024.

\section*{Conflict of Interest declaration}
The authors declare there are no conflicts of interest for this manuscript.

\acknowledgments
The authors would like to thank M. A. Shay for discussions surrounding this work. CLW was supported by UKRI/STFC studentship ST/X508433/1. JPE, HCL and NF acknowledge support from UKRI/STFC grant ST/W001071/1.

%%%%%%%%%%%%%%%%%%%%%%%%%%%%%%%%%%%%%%%%%%%%%%%
% REFERENCES and BIBLIOGRAPHY
%
% \bibliography{<name of your .bib file>} don't specify the file extension
% don't specify bibliographystyle
%
%%%%%%%%%%%%%%%%%%%%%%%%%%%%%%%%%%%%%%%%%%%%%%%

\bibliography{references}

%Reference citation instructions and examples:
%
% Please use ONLY \cite and \citeA for reference citations.
% \cite for parenthetical references
% ...as shown in recent studies (Simpson et al., 2019)
% \citeA for in-text citations
% ...Simpson et al. (2019) have shown...
%
%
%...as shown by \citeA{jskilby}.
%...as shown by \citeA{lewin76}, \citeA{carson86}, \citeA{bartoldy02}, and \citeA{rinaldi03}.
%...has been shown \cite{jskilbye}.
%...has been shown \cite{lewin76,carson86,bartoldy02,rinaldi03}.
%... \cite <i.e.>[]{lewin76,carson86,bartoldy02,rinaldi03}.
%...has been shown by \cite <e.g.,>[and others]{lewin76}.
%
% apacite uses < > for prenotes and [ ] for postnotes
% DO NOT use other cite commands (e.g., \citet, \citep, \citeyear, \nocite, \citealp, etc.).
%

\end{document}